\documentclass[conference]{IEEEtran}
\IEEEoverridecommandlockouts

\usepackage{cite}

\ifCLASSINFOpdf
\else
  \usepackage[dvips]{graphicx}
\fi
\usepackage[cmex10]{amsmath}
\usepackage{array}
\usepackage[tight,footnotesize]{subfigure}
\usepackage{url}
\usepackage{amsthm}
\usepackage{amssymb}
\theoremstyle{plain}

\theoremstyle{remark}

\begin{document}
\title{Pilot Decontamination Through Pilot Sequence Hopping in Massive MIMO Systems}

\author{
\IEEEauthorblockN{
Jesper H. S\o rensen and Elisabeth de Carvalho
}
\IEEEauthorblockA{
Aalborg University, Department of Electronic Systems,
Fredrik Bajers Vej 7, 9220 Aalborg, Denmark
\\
E-mail: \{jhs,edc\}@es.aau.dk
\thanks{The research presented in this paper was supported by the Danish Council for Independent Research (Det Frie Forskningsr{\aa}d) DFF - $1335-00273$}
}
}
\maketitle

\begin{abstract}
%\boldmath
This work concerns wireless cellular networks applying massive multiple-input multiple-output (MIMO) technology. In such a system, the base station in a given cell is equipped with a very large number (hundreds or even thousands) of antennas and serves multiple users. Estimation of the channel from the base station to each user is performed at the base station using an uplink pilot sequence. Such a channel estimation procedure suffers from pilot contamination. Orthogonal pilot sequences are used in a given cell but, due to the shortage of orthogonal sequences, the same pilot sequences must be reused in neighboring cells, causing pilot contamination. The solution presented in this paper suppresses pilot contamination, without the need for coordination among cells. Pilot sequence hopping is performed at each transmission slot, which provides a randomization of the pilot contamination. Using a modified Kalman filter, it is shown that such randomized contamination can be significantly suppressed. Comparisons with conventional estimation methods show that the mean squared error can be lowered as much as an order of magnitude at low mobility.
\end{abstract}
\IEEEpeerreviewmaketitle

\section{Introduction} \label{sec:introduction}
Muliple-input multiple-output (MIMO) technology \cite{edcBook2012} is finding its way into practical systems, like LTE and its successor LTE-Advanced. It is a key component for these systems' ability to improve the spectral efficiency. The success of MIMO technology has motivated research in extending the idea of MIMO to cases with hundreds, or even thousands of antennas, at transmitting and/or receiving side. This is often termed \textit{massive MIMO}. In mobile communication systems, like LTE, the more realistic scenario is to have a massive amount of antennas only at the base station (BS), due to the physical limitations at the user equipment (UE). It has been shown that such a system \cite{marzetta06}, in theory, can eliminate entirely the effect of small-scale fading and thermal noise, when the number of BS antennas goes to infinity. The only remaining impairment is inter-cell interference, caused by imperfect channel state information (CSI), which is a result of non-orthogonality of training pilots used to gather the CSI. This is often referred to as \textit{pilot contamination}. It is considered as one of the major challenges in massive MIMO systems \cite{rusek13}. 

Mitigation of pilot contamination has been the focus of several works recently. These fall into two categories; one with coordination among cells and one without. The first category includes \cite{gesbert13}, where it is utilized that the desired and interfering signals can be distinguished in the channel covariance matrices, as long as the angle-of-arrival spreads of desired and interfering signals do not overlap. A pilot coordination scheme is proposed to help satisfying this condition. The work in \cite{ashikhmin12} utilizes coordination among base stations to share downlink messages. Each BS then performs linear combinations of messages intended for users applying the same pilot sequence. This is shown to eliminate interference when the number of base station antennas goes to infinity.

The category without coordination also includes notable contributions. A multi-cell precoding technique is used in \cite{jose11} with the objective of not only minimizing the mean squared error of the signals of interest within the cell, but also minimizing the interference imposed to other cells. In \cite{ngo12} it is shown that channel estimates can be found as eigenvectors of the covariance matrix of the received signal when the number of base station antennas grows large and the system has ``favorable propagation''. The work in \cite{muller14,muller13,muller13_2,cottatellucci13} is based on examining the eigenvalue distribution of the received signal to identify an interference free subspace on which the signal is projected. It is shown that an interference free subspace can be identified when certain conditions are fulfilled concerning the number of base station antennas, user equipment antennas, channel coherence time and the signal-to-interference ratio.

The major contribution of this paper is a pilot decontamination, which does not require inter-cell coordination, and is able to exploit past pilot signals. It is based on pilot sequence hopping performed within each cell. Pilot sequence hopping means that every user chooses a new pilot sequence in each transmission slot. Consider a user of interest and the effect of the inter-cell pilot contamination when pilot sequence hopping is applied. At each transmission slot, the pilot signal of the user is contaminated by a different set of interfering users. Hence channel estimation at each transmission slot is affected by a different set of interfering channels. If channel estimation is carried out based solely on the pilot sequence of the current slot, then pilot sequence hopping does not bring any gain. The key in our solution is a channel estimation that incorporates multiple time slots so that it can benefit from randomization of the pilot contamination. Recent work utilizing temporal correlation for channel estimation is found in \cite{choi14}, although not in combination with pilot hopping and not with the purpose of mitigating pilot contamination.

Consider the simple example, where the channel of the UE of interest is time-invariant. Its estimation is performed across multiple time slots. Specifically, the resulting channel estimate is the average of the estimates across the time slots. In the averaging process, the contamination signal is averaged out. Note that, if the contamination signal remains constant across the time slots, i.e there is no hopping, this averaging brings no benefit (except an averaging of the receive noise). 

When the channel is time-variant and correlated across time slots, it remains possible to exploit the information about the channel across time slots by an appropriate filtering and benefit from contamination randomization. In this paper, channel estimation across multiple time slots is performed using a modified version of the Kalman filter, which is capable of tracking the channel and the channel correlation. The level of contamination suppression depends on the channel correlation between slots of the UE of interest as well as the contaminators. In LTE, channel correlation between time slots is large even at medium-high speeds, making the proposed solution very efficient.

The remainder of this paper is organized as follows. Section \ref{sec:sysmodel} presents the applied system model and the problem of pilot contamination. The proposed solution is described in section \ref{sec:scheme} and evaluated and compared to existing solutions in section \ref{sec:results}. Finally, conclusions are drawn in section \ref{sec:conclusions}.

%The solution presented in this paper belongs to the category without coordination. Compared to the mentioned work it carries novelty by utilizing information in past pilot signals, which exists in practical systems with pilot schedules designed for worst-case, but rare, mobility scenarios. When combined with a pilot sequence hopping scheme, it is shown to suppress a significant part of the contamination.
\section{System Model}\label{sec:sysmodel}
In this work we denote scalars in lower case, vectors in bold lower case and matrices in bold upper case. A superscript ``$T$'' denotes the transpose and a superscript ``$H$'' denotes the conjugate transpose.

This work treats a cellular system consisting of $L$ cells with $K$ users in each cell. A massive MIMO scenario is considered, where the BS has $M$ antennas and the UE has a single antenna. We restrict our attention to the channel estimation performed in a single cell, which we term ``the cell of interest'' and assign the index ``$0$''. The channel between the BS in the cell of interest and the $k$'th user in the $\ell$'th cell is denoted $\pmb{h}^{k \ell}=\left[h^{k \ell}(1)\hspace{0.1cm}h^{k \ell}(2) \hdots h^{k \ell}(M)\right]$, where the individual channel coefficients are complex scalars. Note that for $\ell>0$, $\pmb{h}^{k \ell}$ refers to a channel between the BS of interest and a UE connected to a different base station. We furthermore restrict our attention to the estimation of a single channel coefficient, hence a channel is denoted as the complex scalar $h^{k\ell}$. The work easily extends to vector estimations, in which case spatial correlation can be exploited for improved performance. A rich scattering environment is assumed, such that $h^{k\ell}$ can be modeled using Clarke's model \cite{clarke68}, hence

\begin{align}
h^{k\ell} = \frac{1}{\sqrt{N_s}}\sum_{m=1}^{N_s} \mathrm{e}^{j2\pi f_d t \cos\alpha_m+\phi_m},
\end{align}

\noindent where $N_s$ is the number of scatterers, $f_d$ is the maximum Doppler shift, $\alpha_m$ and $\phi_m$ is the angle of arrival and initial phase, respectively, of the wave from the $m$'th scatterer. Both $\alpha_m$ and $\phi_m$ are i.i.d. in the interval $[-\pi,\pi)$ and $f_d=\frac{v}{c}f_c$, where $v$ is the speed of the UE, $c$ is the speed of light and $f_c$ is the carrier frequency.

\begin{figure}[ht]
 \centering
 \includegraphics[width=1.0\columnwidth]{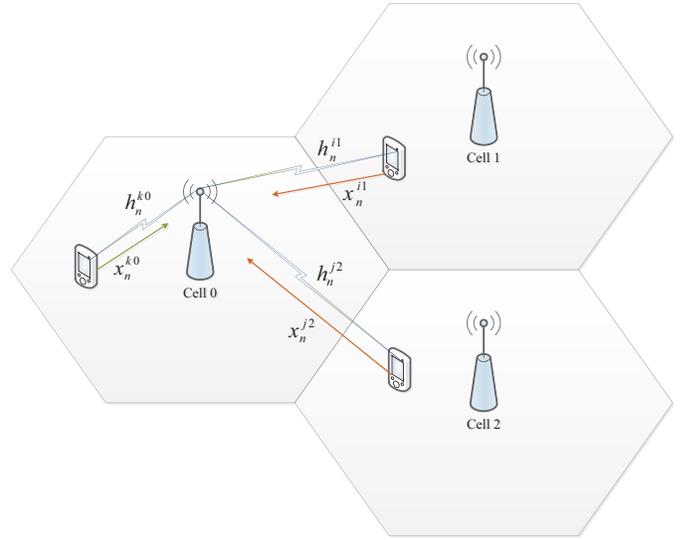}
 \caption{A cellular system with three cells. Cell $0$ is of interest and the neighboring cells will potentially cause interference (red arrows).}
 \label{fig:hexagrid}
\end{figure}

\begin{figure}[ht]
 \centering
 \includegraphics[width=1\columnwidth]{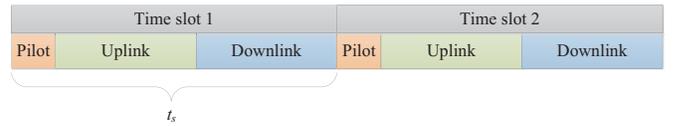}
 \caption{Scheduling example.}
 \label{fig:schedule}
\end{figure}

In a massive MIMO system, collection of channel state information (CSI) is performed using uplink pilot training. The CSI achieved this way is utilized in both downlink and uplink transmissions based on the channel reciprocity assumption. We define a pilot training period followed by an uplink and a downlink transmission period as a time slot. See Fig. \ref{fig:schedule} for an example of a transmission schedule with two time slots. During the $n$'th pilot training period, the $k$'th user in the $\ell$'th cell transmits a pilot sequence $\pmb{x}^{k\ell}_n=\left[x^{k\ell}_n(1)\hspace{0.1cm}x^{k\ell}_n(2) \hdots x^{k\ell}_n(\tau)\right]^T$, where $\tau$ is the pilot sequence length. Ideally, all pilot sequences in the entire system are orthogonal, in order to avoid interference. However, this would require pilot sequences of at least length $L \cdot K$, which in most practical systems is not feasible. Instead, orthogonality within each cell only is ensured, i.e. $\tau=K$, thereby dealing with the potentially strongest sources of interference. As a result, all cells use the same set of pilots, potentially causing interference from neighboring cells. This is referred to as pilot contamination. We define the contaminating set, $\mathcal{C}_n^{k\ell}$, as the set of all pairs $i,j$, which identify all UEs applying the same pilot sequence in the $n$'th time slot as the $k$'th user in the $\ell$'th cell. Hence, $\pmb{x}^{ij}_n = \pmb{x}^{k\ell}_n$ $\forall$ $i,j\in\mathcal{C}_n^{k\ell}$.

The pilot signal received by the BS of interest, concerning the $k$'th user in the $n$'th time slot can be expressed as

\begin{align}\label{eq:obs}
\pmb{y}^{k0}_n = h^{k0}_n \pmb{x}^{k0}_n + \sum_{i,j\in\mathcal{C}_n^{k0}} h^{ij}_n \pmb{x}^{ij}_n + \pmb{z}^{k0}_n,
\end{align}

\noindent where $\pmb{z}^{k0}_n=\left[z^{k0}_n(1)\hspace{0.1cm}z^{k0}_n(2) \hdots z^{k0}_n(\tau)\right]^T$ and $z^{k0}_n(j)$ are circularly symmetric Gaussian random variables with zero mean and unit variance for all $j$. Here, only signals leading to contamination are included in the sum term, since any $h^{ij}_n \pmb{x}^{ij}_n$ $\forall$ $i,j\notin\mathcal{C}_n^{k\ell}$ are removed when correlating with the applied pilot sequence. Hence, all contributions from the sum term are undesirable and will contaminate the CSI. Without loss of generality, we focus on the channel estimation for a single user in a single cell. Hence, in the remainder of the paper, we omit the superscript $k$ for ease of notation.

%h(k)=sum(exp(1i*(2*pi*fd*t(k)*cos(alpha)+phi)),2)/sqrt(ns);
%y(:,k)=y(:,k)+h(k)*x+sqrt(Pn)*(randn(tau,1)+1i*randn(tau,1))/sqrt(2);
%for l=1:nc
%	hc(l,k)=sum(exp(1i*(2*pi*fd*t(k)*cos(alpha_c(:,l,k))+phi_c(:,l,k))))/sqrt(ns);
%	y(:,k)=y(:,k)+(sqrt(Pc)*hc(l,k).'*x.').';
%end

%During uplink pilot training 

%\[x_{k\ell n1} x_{k\ell n2} \cdots x_{k\ell n\tau}\]

%, see Fig. \ref{fig:hexagrid}
\section{Pilot Decontamination}\label{sec:scheme}
The solution to pilot contamination proposed in this work consists of two components:
\begin{enumerate}
\item \textbf{Pilot sequence hopping:} This component refers to random shuffling of the pilots applied within a cell. This shuffle occurs between every time slot. The purpose of this component is to \textit{decorrelate} the contaminating signals. When pilots are shuffled, the set of contaminating users will be replaced by a new set, whose channel coefficients are uncorrelated with those of the previous set.
\item \textbf{Kalman filtering:} The autocorrelation of the channel coefficient of the user of interest is high at low mobility. This means that information about the value of the current channel coefficient exists not only in the most recent pilot signal, but also in past pilot signals. This can be extracted using a filter. For this purpose a Kalman filter is desirable due to its recursive structure, which provides low complexity, yet optimal performance. Additionally, since the contaminating signals have been decorrelated, the Kalman filter will suppress the impact of these signals, leading to pilot \textit{decontamination}.
\end{enumerate}

\subsection{Pilot Sequence Hopping}
\begin{figure}[ht]
 \centering
 \includegraphics[width=1\columnwidth]{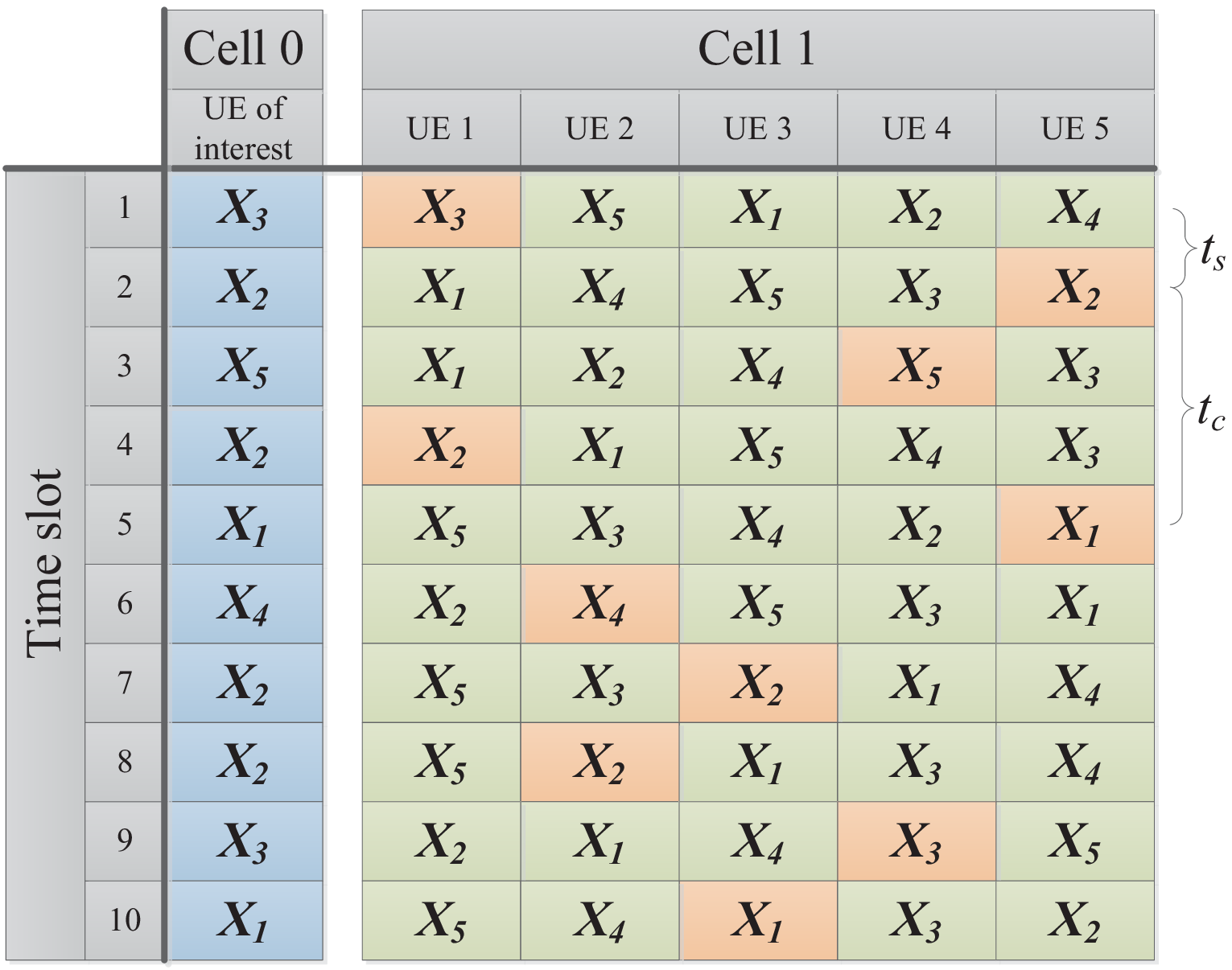}
 \caption{An example of a random pilot schedule for the UE of interest and potential contaminators in a neighboring cell. Green boxes represent pilots, which are orthogonal to the pilot from the UE of interest. Red boxes represent contamination and $\pmb{x}_i$ denotes a pilot sequence.}
 \label{fig:hopping}
\end{figure}

Pilot sequence hopping is a technique where the UEs randomly switch to a new pilot sequence in between time slots. This must be coordinated with the BS, which in practice can be realized by letting the BS send a seed for a pseudorandom number generator to each UE. Random pilot sequence hopping is illustrated in Fig. \ref{fig:hopping} in the case of $\tau=K=5$. Note how the identity of the contaminator changes between time slots, as opposed to a fixed pilot sequence schedule, where the contaminator remains the same UE. Consequently, the undesirable part of the pilot signal, i.e. the sum term in \eqref{eq:obs}, varies rapidly between time slots compared to the variation caused by the mobility of a single contaminator in a fixed schedule. In fact, the impact of pilot sequence hopping, from a contamination perspective, can be viewed as a dramatic increase of the mobility of the contaminator. This in turn leads to a lowered autocorrelation, or decorrelation, in the contaminating signal, which is the motivation behind performing pilot sequence hopping.

The level of decorrelation is related to the time between two instances, where the same user acts as a contaminator. We refer to this as the collision distance, and we denote it $t_c$, see Fig. \ref{fig:hopping}. Note that in the case of a fixed pilot schedule, $t_c=1$. The goal of pilot sequence hopping is to maximize $t_c$, either in an expected sense or maxmin sense, i.e. maximization of the minimum value. The latter can be pursued through a minimal level of coordination of pilot sequence schedules among neighboring cells. However, this work is strictly restricted to a framework with no inter-cell coordination, hence, we focus on the expected value of $t_c$. If pilot sequence hopping is performed at random and $\tau=K$, then $t_c$ follows a geometric distribution, such that

\begin{align}
P(t_c=d) &= (1-p)^{d-1}p, \hspace{1cm} d=1,2,\hdots, \notag \\
p &= \frac{1}{K},
\end{align}

\noindent where $P(t_c=d)$ is the probability that the collision distance is $d$ and $p$ is the probability of a given UE being the next contaminator. The expected value of $t_c$, $\mathbb{E}\left[t_c\right]$, is then found as

\begin{align}
\mathbb{E}\left[t_c\right] &= \sum_{d=1}^{\infty} d(1-p)^{d-1}p \notag \\
&= \sum_{d=1}^{\infty} d\left(\frac{K-1}{K}\right)^{d-1}\frac{1}{K} \notag \\
&=K. \label{eq:tc}
\end{align}

\noindent Hence, the expected collision distance increases with the number of users/pilots per cell, which follows intuition.

\textbf{Example:} To help the understanding of the benefit from pilot sequence hopping, consider the ideal case of a constant channel between BS and UE of interest and a single contaminating neighboring cell. Noise is disregarded in this example, since attention is on decontamination. Moreover, we assume an infinite amount of orthogonal pilot sequences and an infinite amount of users per cell, such that $\tau=K=\infty$ and $\mathbb{E}\left[t_c\right]=\infty$, which means contaminating signals in all time slots are independent. For simplicity, we assume $\pmb{x}_n^H \pmb{x}_n =1$, such that the estimate in time slot $n$ is

\begin{align}
\hat{h}_n=h+h_n',
\end{align}

\noindent where $h_n'$ is the channel of the contaminator in time slot $n$. Now consider a new estimator, $\bar{\hat{h}}_n$, which is the average of all estimates until time slot $n$. Hence, we have

\begin{align}
\bar{\hat{h}}_n&=h+ \frac{1}{n}\sum_{i=1}^n h_i'.
\end{align}

In this case, the error in the estimate is solely composed of the average of the contaminating signals, which are independent and have variance $\sigma_c^2$. Hence, the variance of the estimation error is $\frac{\sigma_c^2}{n}$. If pilot sequence hopping had not been performed, the variance of the estimation error had remained $\sigma_c^2$, since $h_n'$ would be constant. Note that the MSE goes towards zero for $n\to\infty$, when pilot sequence hopping is performed. This is a result of the fact that a pilot signal in the infinite past carries as much information about the current channel as the most recent pilot signal, in the ideal example of a constant channel. Note also that for finite $\tau$ (and $K$) and thereby finite $\mathbb{E}\left[t_c\right]$, the variance of the estimation error is lower bounded by $\frac{\sigma_c^2}{K}$, since only a maximum of $K$ independent estimates can be achieved. In a more practical example with a time-varying channel, the amount of information carried in a pilot signal decays over time. It is, however, still possible to extract such information using appropriate filtering techniques. For this purpose we have chosen a modified version of the Kalman filter, which is described next.

%amplitudes are Rayleigh distributed and independent. Hence, the mean squared error (MSE) of the estimate is $\frac{4-\pi}{2n}\sigma_c^2$, where $\sigma_c^2$ is the power of a contaminating signal.

\subsection{Modified Kalman Filter}\label{sec:mod}
A conventional Kalman filter can be used to track the state, $\pmb{b}_n$, of a system based on observations, $\pmb{y}_n$, where

\begin{align}\label{eq:mes}
\pmb{y}_n &= \pmb{C}_n \pmb{b}_n + \pmb{d}_n,
\end{align}

\noindent and $\pmb{C}_n$ is the measurement matrix of the system and $\pmb{d}_n$ is measurement noise. Moreover, the evolution of the system state must follow

\begin{align}\label{eq:ar}
\pmb{b}_n &= \pmb{A}_n \pmb{b}_{n-1} + \pmb{v}_n,
\end{align}

\noindent where $\pmb{A}_n$ is the state transition matrix and $\pmb{v}_n$ is the process noise. In a conventional application of the Kalman filter, $\pmb{A}_n$ is assumed constant and known.

The problem of estimating a time-varying channel based on pilot signals, also termed channel tracking, can be solved using the Kalman filter. The observations as expressed in \eqref{eq:obs} follow the linear model in \eqref{eq:mes}, where the observation matrix is the transmitted pilot sequence and the tracked state is the channel coefficient. The evolution of the channel coefficient as expressed by Clarke's model does not follow the model in \eqref{eq:ar}. However, it can be transformed into an autoregressive (AR) model with a finite number of coefficients, which follows the form of \eqref{eq:ar}. If the instantaneous velocity of the user of interest, and thereby the autocorrelation function, are known, the AR coefficients can be found using the Yule-Walker equations \cite{kay81}. However, this cannot be assumed in our case, hence the AR coefficients must be tracked along with the channel state. For this purpose, we must modify the conventional Kalman filter to include an AR model tracker. A $1$\textsuperscript{st} order AR model is applied, since experiments tell us this adequately captures the autocorrelation of the system. Therefore, only a single AR coefficient, $a_n$, must be tracked.

First we state the conventional Kalman filter \cite{haykin13} in our context, where the AR coefficient is assumed known.\\

\noindent For all $n$:
\begin{align}
\pmb{e}_n &= \pmb{y}_n-\pmb{x}_n a_{n-1} \hat{h}_{n-1}, \label{e} \\
\pmb{R}_n &= \pmb{x}_n p_n \pmb{x}_n^H + \sigma^2_n \pmb{I}_{\tau} + \sigma^2_c \pmb{x}_n \pmb{x}_n^H, \\
\pmb{k}_n &= p_n \pmb{x}_n^H \pmb{R}_n^{-1}, \label{k}\\
\hat{h}_n &= a_n \hat{h}_{n-1} + \pmb{k}_n \pmb{e}_n, \label{hath}\\
p_{n+1} &= a_n^2 (1-\pmb{k}_n \pmb{x}_n) p_n + (1-a_n^2), \label{p}
\end{align}

\noindent where $\sigma^2_n$ and $\sigma^2_c$ are noise power and total contamination power (average over time), respectively, which are both assumed known, $\pmb{I}_{\tau}$ is the $\tau\times\tau$ identity matrix and $\hat{h}_n$ is the estimate of $h_n$.

For the tracking of the AR coefficient, an approach similar to the one in \cite{han04} is taken. In \cite{han04} the inclusion of an AR coefficient tracker is presented for a Kalman predictor, i.e. a filter with the purpose of predicting the channel, $h_n$, based on all observations until $\pmb{y}_{n-1}$. In this work, we extend this approach to take all observations until $\pmb{y}_n$ into account.

The approach is based on calculating the partial derivative with respect to $a_n$ of the cost function, the mean squared error (MSE), and using this to adjust $a_n$ in the direction of decreasing MSE. The partial derivative of the MSE is

\begin{align}
\nabla_n &= \frac{\partial}{\partial a_n} \mathbb{E}\left[|\pmb{e}_n|^2\right] \notag \\
         &= -(q_{n-1}^H a_{n-1} \pmb{x}_n^H + \hat{h}_{n-1}^H \pmb{x}_n^H) \pmb{e}_n,
\end{align} 

\noindent where $q_{n}=\frac{\partial \hat{h}_{n}}{\partial a_n}$ and is found by differentiating \eqref{hath} with respect to $a_n$, such that

\begin{align}
q_n &= (1-\pmb{k}_n \pmb{x}_n) (a_n q_{n-1} + \hat{h}_{n-1}) + \pmb{m}_n \pmb{e}_n.
\end{align}

\noindent Here, $\pmb{m}_n=\frac{\partial \pmb{k}_n}{\partial a_n}$, which is found by differentiating \eqref{k} with respect to $a_n$, hence

\begin{align}
\pmb{m}_n &= (1-\pmb{k}_n \pmb{x}_n) s_n \pmb{x}_n^H \pmb{R}_n^{-1}.
\end{align}

\noindent Finally, we introduced $s_n=\frac{\partial p_n}{\partial a_n}$, which is a differentiation of \eqref{p} with respect to $a_n$, giving us

\begin{align}
s_{n+1} &= a_n^2 (1-\pmb{k}_n \pmb{x}_n) s_n (1-\pmb{x}_n^H \pmb{k}_n^H) - 2 a_n \pmb{k}_n \pmb{x}_n p_n.
\end{align}

\noindent Using $\nabla_n$, we can adjust $a_n$ as follows

\begin{align}
a_n &= [a_{n-1} - \mu [\nabla_n]_{-\nu}^{+\nu}]_0^1,
\end{align}

\noindent where $\mu$ is a parameter adjusting the convergence speed and the brackets denote truncations. The inner truncation involving $\nu$ is to avoid dramatic adjustments in situations with a high slope and the outer truncation is to obey $0\le a_n \le 1$. The need for $\nu$ will be explained in section \ref{sec:results}.

We can now state the modified Kalman filtering algorithm including an AR coefficient tracker:\\

\noindent For all $n$:
\begin{align}
\pmb{e}_n &= \pmb{y}_n-\pmb{x}_n a_{n-1} \hat{h}_{n-1}, \notag \\
\pmb{R}_n &= \pmb{x}_n p_n \pmb{x}_n^H + \sigma^2_n \pmb{I}_{\tau} + \sigma^2_c \pmb{x}_n \pmb{x}_n^H, \notag \\
\nabla_n &= -(q_{n-1}^H a_{n-1} \pmb{x}_n^H + \hat{h}_{n-1}^H \pmb{x}_n^H) \pmb{e}_n, \notag \\
a_n &= [a_{n-1} - \mu [\nabla_n]_{-\nu}^{+\nu}]_0^1, \notag \\
\pmb{k}_n &= p_n \pmb{x}_n^H \pmb{R}_n^{-1}, \notag \\
\hat{h}_n &= a_n \hat{h}_{n-1} + \pmb{k}_n \pmb{e}_n, \notag \\
\pmb{m}_n &= (1-\pmb{k}_n \pmb{x}_n) s_n \pmb{x}_n^H \pmb{R}_n^{-1}, \notag \\
q_n &= (1-\pmb{k}_n \pmb{x}_n) (a_n q_{n-1} + \hat{h}_{n-1}) + \pmb{m}_n \pmb{e}_n, \notag \\
p_{n+1} &= a_n^2 (1-\pmb{k}_n \pmb{x}_n) p_n + (1-a_n^2), \notag \\
s_{n+1} &= a_n^2 (1-\pmb{k}_n \pmb{x}_n) s_n (1-\pmb{x}_n^H \pmb{k}_n^H) - 2 a_n \pmb{k}_n \pmb{x}_n p_n.
\end{align}

%e=y(:,k)-x*F(k)*h_est(k);
%r=(x*K(k)*x'+Pn*eye(tau)+nc*Pc*(x*x'))^-1;
%grad=real((q(k)'*F(k)*x'+h_est(k)'*x')*(y(:,k)-x*F(k)*h_est(k)));
%grad=max(-nu,min(nu,grad));
%F(k+1)=max(10^-6,min(1-10^-6,F(k)+mu*grad));
%G=K(k)*x'*r;
%m=S(k)*x'*r-G*x*S(k)*x'*r;
%q(k+1)=-G*x*F(k+1)*q(k)-G*x*h_est(k)+F(k+1)*q(k)+h_est(k)+m*e;
%S(k+1)=F(k+1)^2*(1-G*x)*S(k)*(1-x'*G')-2*F(k+1)+2*F(k+1)*(1-G*x)*K(k);
%h_est(k+1)=F(k+1)*h_est(k)+G*e;
%K(k+1)=F(k+1)*(K(k)-G*x*K(k))*F(k+1).'+(1-F(k+1)^2);
%\input{analysis}
\section{Numerical Results} \label{sec:results}
The proposed scheme (Estimator) has been simulated and compared to the scheme from \cite{han04} (Predictor) and the conventional solutions of least squares (LS) estimation and minimum mean squared error (MMSE) estimation based on a single time slot. The expressions for the LS and MMSE estimators are given in \eqref{eq:ls} and \eqref{eq:mmse}, respectively. An overview of the parameters, which are common for all simulations, is given in Table \ref{tab:params}. The choice of $\mu$ is based on experiments showing that this is a good compromise between convergence speed and robustness towards variance. Throughout all simulations, we assume that all users have equal and constant mobility. Moreover, we assume that contaminating signals have zero autocorrelation between time slots, which is justified by the choice of $K=96$, such that $\mathbb{E}\left[t_c\right]=96$, cf. \eqref{eq:tc}.

\begin{align}
\hat{h}_n^{ls}&=\left(\pmb{x}_n^H \pmb{x}_n\right)^{-1} \pmb{x}_n^H \pmb{y}_n, \label{eq:ls}\\
\hat{h}_n^{mmse}&=\pmb{x}_n^H\left(\pmb{x}_n \pmb{x}_n^H+\sigma_n^2 \pmb{I}_{\tau} + \sigma_c^2 \pmb{x}_n \pmb{x}_n^H\right)^{-1}\pmb{y}_n. \label{eq:mmse}
\end{align}

%G=x'*(x*x'+Pn*eye(tau)+nc*Pc*(x*x'))^-1;
%            h_est(k)=G*y(:,k);

\begin{table}[ht]
\centering
	\caption{Simulation parameters}
	\label{tab:params}
\begin{tabular}{|c|c|c|} \hline
  Parameter & Value & Description \\ \hline \hline
  \rule{0pt}{3ex} $\sigma_n^2$ & $0.2$ & Noise variance\\ \hline
	\rule{0pt}{3ex} $L$ & $7$ & Number of cells\\ \hline
	\rule{0pt}{3ex} $K$ & $96$ & Users per cell\\ \hline
	\rule{0pt}{3ex} $\tau$ & $96$ & Pilot length \\ \hline
  \rule{0pt}{3ex} $\mu$        & $10^{-5}$ & Convergence speed\\ \hline
	\rule{0pt}{3ex} $\nu$        & $100$ & Derivative cap\\ \hline
	\rule{0pt}{3ex} $f_c$        & $1.8\hspace{0.1cm}$GHz & Carrier frequency\\ \hline
	\rule{0pt}{3ex} $N_s$ & $20$ & Number of scatterers \\ \hline
	\rule{0pt}{3ex} $t_s$ & $0.5\hspace{0.1cm}$ms & Time between pilots \\ \hline
	\rule{0pt}{3ex} $a_0$ & $0.5$ & Initial AR coefficient \\ \hline
	\rule{0pt}{3ex} $\hat{h}_0$ & $0$ & Initial estimate \\ \hline
	\rule{0pt}{3ex} $q_0$ & $0$ & Initial differentiated estimate \\ \hline
	\rule{0pt}{3ex} $p_1$ & $0$ & Initial error covariance \\ \hline
	\rule{0pt}{3ex} $s_1$ & $0$ & Initial differentiated error covariance \\ \hline
\end{tabular}
\end{table}

\begin{figure}[ht]
 \centering
 \includegraphics[width=1\columnwidth]{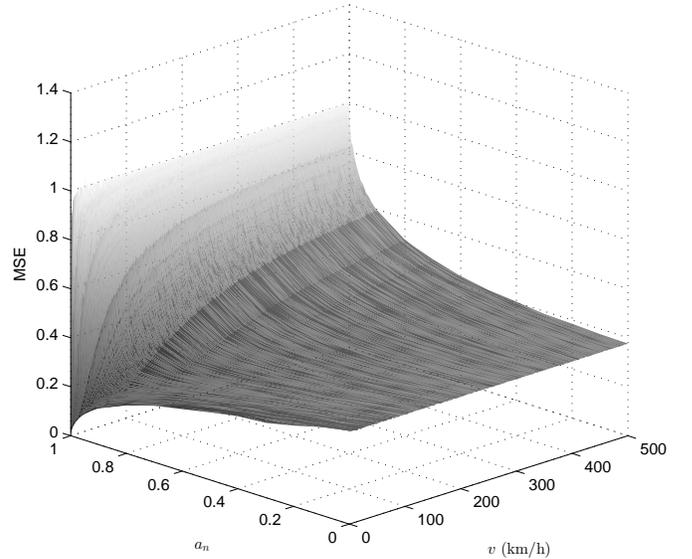}
 \caption{MSE as a function of the autoregressive model coefficient and the user mobility.}
 \label{fig:surf}
\end{figure}

Initially, results are shown for the conventional Kalman filter expressed in equations \eqref{e} through \eqref{p}. MSE as a function of the user mobility, $v$, and the AR coefficient, $a_n$, is shown in Fig. \ref{fig:surf}. From this figure, it is evident how important it is to have an accurate AR model, which suits the current mobility of the UE of interest. This stresses the need for the modification of the Kalman filter, as proposed in section \ref{sec:mod}. Moreover, it is seen that the derivative of the MSE with respect $a_n$ may attain very high values at low $a_n$. This can cause undesirably high variance in the estimate of the optimal $a_n$, which motivates the use of a derivative cap, $\nu$.

\begin{figure}[ht]
 \centering
 \includegraphics[width=1\columnwidth]{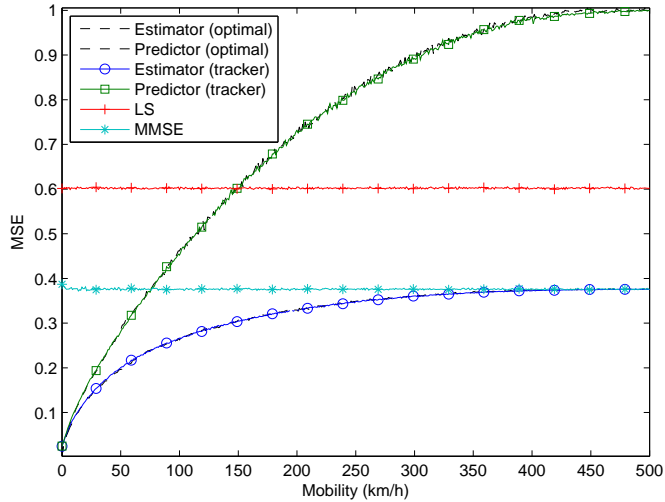}
 \caption{Comparison between the proposed scheme and conventional solutions with respect to means squared error as a function of mobility.}
 \label{fig:results_mob}
\end{figure}

Fig. \ref{fig:results_mob} shows a comparison of the simulated estimators with respect to MSE as a function of user mobility when $\sigma_c^2=0.6$. For both the predictor and the scheme proposed in this work, results where the optimal value of $a_n$ is assumed to be known, have been included. This highlights the performance of the tracker. It is evident that the tracker provides a very good estimate of the optimal AR coefficient. Moreover, it is seen that the proposed scheme outperforms LS and MMSE and performs as well as the predictor at low mobility. At high mobility, the proposed scheme outperforms LS and the predictor, while matching the performance of MMSE.

\begin{figure}[ht]
 \centering
 \includegraphics[width=1\columnwidth]{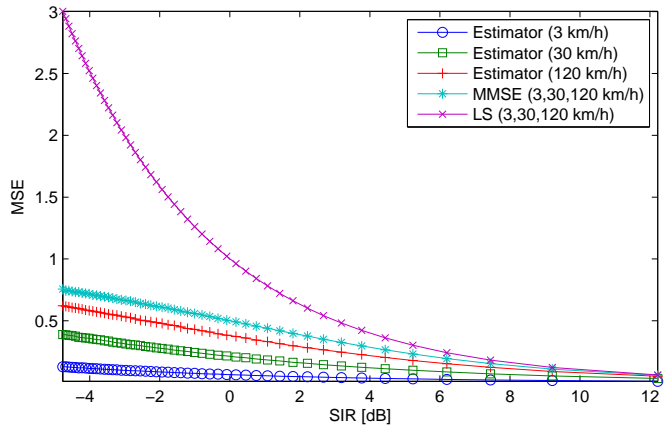}
 \caption{Comparison between the proposed scheme and conventional solutions with respect to means squared error as a function of the signal-to-interference ratio.}
 \label{fig:results_sir}
\end{figure}

A different perspective is given in Fig. \ref{fig:results_sir}. Here the MSE is plotted as a function of the signal-to-interference ratio (SIR), at typical mobility levels as defined by 3GPP \cite{3gpp.25.996}. This figure shows how the proposed scheme is able to suppress even very strong contamination at typical mobility.

%3,30,120 km/h cite=3GPP TR 25.996 
\section{Conclusions}\label{sec:conclusions}
We have presented a solution to pilot contamination in channel estimation, which is a major challenge in massive MIMO systems. It is based on a combination of a pilot sequence hopping scheme and a modified Kalman filter. The pilot sequence hopping scheme involves random shuffling of the assigned pilot sequences within a cell, which ensures decorrelation in the time dimension of the contaminating signals. This is essential, since it enables subsequent filtering to suppress the contamination. For this filtering, the Kalman filter has been chosen, due to its ability to track a time-varying state. However, a conventional Kalman filter is not able to adapt to changes in the underlying model, which is necessary when users have unknown and varying levels of mobility. For this problem we have presented a modified Kalman filter, which can adapt the underlying model based on a minimization of the mean squared error.

Numerical evaluations show that the proposed solution can suppress a significant portion of the contamination at low and moderate levels of mobility. Even at high mobility, i.e. car speeds of $100$ to $130$ km/h, the proposed solution can provide a noticeable gain over conventional estimation methods.

\bibliographystyle{ieeetr}
\bibliography{bibliography}

\end{document}